\documentclass[twoside]{article}
\usepackage[dvips]{graphicx}
\usepackage{fleqn,espcrc2}
\usepackage{pstricks}
\usepackage{pst-node}
\usepackage[figuresright]{rotating}

\newcommand{\be}{\begin{equation}}
\newcommand{\ee}{\end{equation}}
\newcommand{\bc}{\begin{center}}
\newcommand{\ec}{\end{center}}
\newcommand{\bes}{\begin{equation*}}
\newcommand{\ees}{\end{equation*}}
\newcommand{\beqn}{\begin{eqnarray}}
\newcommand{\eeqn}{\end{eqnarray}}
\newcommand{\beqns}{\begin{eqnarray*}}
\newcommand{\eeqns}{\end{eqnarray*}}

\newcommand{\err}[2]{\raisebox{0.08em}{\scriptsize {$\;\begin{array}{@{}l@{}}
			  \plus\makebox[2.05em][r]{#1} \\[-0.12em] 
			  \minus\makebox[2.05em][r]{#2} 
			\end{array}$}}}

\newcommand{\plus}{\makebox[2pt][c]{$+$}}
\newcommand{\minus}{\makebox[2pt][c]{$-$}}

\newcommand{\AmS}{{\protect\the\textfont2
  A\kern-.1667em\lower.5ex\hbox{M}\kern-.125emS}}

\title{Hyperons, Charm and Beauty Hadrons:\\ Conclusion and Outlook}

\author{Jos\'e Bernab\'eu\address{Dep. de F\'\i sica Te\`orica, Univ. de
Val\`encia,\\  
Dr. Moliner 50, E-46100, Burjassot, Val\`encia, Spain}}
\begin{document}

\begin{abstract}
In this concluding talk, the advances in the Flavour Problem studies are 
discussed, following the structure of the presentations in the Conference. 
The subjects touched are organized as follows: Baryons, K-physics, Charm and 
Beauty production, Charm and Beauty decays, B-Mixing and CP-Violation, 
Heavy Quarkonium.
\vspace{1pc}
\end{abstract}

\maketitle

\section{Introduction.}
The subjects presented in the Conference \cite{ref0} have in common their contribution to 
the understanding of the Flavour Problem "from below", i.e., from detailed 
studies of the structures, regularities and differences among the flavoured 
hadrons. In this edition, many new interesting results have been presented 
and my discussion will be necessarily limited in scope. I apologize for 
the omissions or simplifications in the conclusions given here.

The quarks carry (among other properties) the flavour quantum number conserved 
by strong and electroweak neutral current interactions to leading order. They 
are organized in three families which appear as replicas. Besides the anthropic
statement that three families is the minimum number able to build a Universe 
with the prospect of being understood by humans through science, we do not 
have still an explanation for the mistery of this replication. Except for weak
charged current interactions, the other fundamental forces are unable to connect
the families each other. In the Standard Model, the CKM Mixing Matrix 
gives account of this problem according to the scheme in Fig.~1.
\begin{figure}[!h]
\includegraphics[bb=2.5cm 5cm 17.5cm 18cm,scale=0.45]{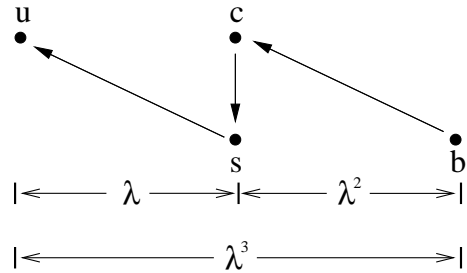}	
\vskip -2cm \caption{CKM Mixing Matrix scheme.}
\end{figure}
The second mistery in the Flavour Problem is the hierarchy of mixings, 
with intensities of order $\lambda$, $\lambda^{2}$, $\lambda^{3}$ for the 
transitions shown in the Fig.~1.
In the step from quarks to hadrons, however, the different quark masses 
provide an essential difference between the structure of light hadrons 
(u, d and s) and that of heavy systems (c, b and t). The plan of this 
contribution is as follows. In Section 2, we discuss Baryons, with some 
emphasis on Hyperons. Section 3 is devoted to K-physics, with the highlight of 
the last two years: KTeV and NA48 confirm the $3.5\, \sigma$ NA31 result of 
direct CP-violation. In Section 4, charm and beauty production, the study of the 
fragmentation function provides the link between quarks and hadrons. 
Section 5 discusses charm and beauty decays, including semileptonic, purely 
leptonic, hadronic and rare decays. The problem of B (and D) Mixing and 
CP Violation is presented in Section 6, with the novel results of BaBar at 
PEP-II and Belle at KEK B. Heavy Quarkonium is discussed in Section 7. 
Finally, Section 8 gives some Outlook.

\section{Baryons.}
The study of semileptonic decays of heavy baryons has been addressed \cite{ref1} 
in a consistent quark model describing baryons. A single set of parameters is 
used for the whole spectra and the resulting structure is tested with decays. 
The investigation of inclusive semileptonic $\Lambda_{b}$ decays can provide 
information on the CKM matrix elements $V_{cb}$ and $V_{ub}$, as well as on 
the structure of $\Lambda_{b}$. 
The  approach is that of a potential model with physical values of the 
couplings. The sum over final hadronic states is treated by means of duality. 
The predicted value of the inclusive semileptonic widths of  $\Lambda_{c}$ and 
$\Lambda_{b}$ are confronted to the experiment \cite{ref2} in Table \ref{tabl1}, 
where the branching ratios are given.

\begin{table}[t]
\parbox{7.5cm}{\caption{Comparison between the predicted and experimental
values of the inclusive semileptonic widths of $\Lambda_{c}$ and 
$\Lambda_{b}$.}\label{tabl1}}\\
\vskip 0.25cm

\begin{center}
\begin{tabular}{ccc}
\hline 
BR$_{SL}$(\%)&Model&Experiment\\
\hline 
\hline 
$\Lambda_{c}$&$5.5$&$4.5\pm 1.7$\\
$\Lambda_{b}$&$10.7$&$9\err{3.1}{3.8}$\\
\hline 
\hline 
\end{tabular}
\end{center}
\end{table}

The exclusive/inclusive ratio ${\rm R}_{E}$ of semileptonic 
$\Lambda_{b}\rightarrow \Lambda_{c}$ decay has been compared with the 
corresponding ratio for the meson $B\rightarrow D\, +\, D^{*}$, with the 
conclusion \cite{ref3} that it should be larger. One needs the slope parameter 
$\rho_{B}^{2}$ of the Isgur-Wise form factor $F_{B}(\omega)$ in 
$\Lambda_{b}\rightarrow \Lambda_{c}\, l\, \nu$
\be\label{eq1}
F_{B}(\omega)\, =\, 1\, -\, \rho^{2}_{B}\, (\omega-1)\, +\, c\, (\omega-1)^{2}\, +\, \cdots
\ee
and an upper bound is taken from the spectator quark model limit
\be
\rho^{2}_{B}\, \leq\, 2\, \rho^{2}_{M}\, -\, \frac{1}{2}
\ee
with the experimental value \cite{ref2} $\rho^{2}_{M}\, =\, 0.70\pm 0.10$. 
From QCD sum rules, $0.65\leq \rho^{2}_{B}\leq 0.85$, and one finds a ratio
\[
0.81\, \leq\, R_{E}({\rm baryon})\, \leq\, 0.92
\]
to be compared with $R_{E}({\rm meson})=66\%$.

The exclusive process $\Lambda_{b}\rightarrow \Lambda_{c}\, l\, \nu_{l}$ has 
been experimentally searched by DELPHI Collaboration \cite{ref4}, with the 
analysis addressed to measure the slope parameter $\rho^{2}_{B}$ of the form 
factor (\ref{eq1}). With appropiate cuts in $p_{l}$ and $p_{\bot}$, 
the invariant masses $M(\Lambda_{c}\, e)$, $M(\Lambda_{c}\, \mu)$, a candidate 
is taken as the sign of $l$ opposite to that of $\Lambda_{c}$. They find 
$57\pm 8$ events and the measure of the slope parameter gives
\be
\rho^{2}_{B}\, =\, 1.6\, \pm\, 0.6\, ({\rm sta})\, \pm\, 0.6\, ({\rm syst})
\ee
when the absolute event rate is included in the fit.

A review on the b hadron lifetimes was presented \cite{ref5} by Wasserbaech, 
ALEPH Coll. Recent measurements from LEP, SLD and CDF indicate that we have 
still a problem with the lifetime of $\Lambda_{b}$. From the theoretical side, 
the measurement of the individual lifetimes of $B^{+}$, $B_{d}$, $B_{s}$ and
$\Lambda_{b}$ yields information about nonspectator mechanisms. The experimental
results for the lifetime ratios are given in Table \ref{tabl2}, together with 
the theoretical predictions from QCD-based Heavy Quark expansions.

\begin{table}[t]
\parbox{7.5cm}{\caption{Comparison between the predicted and experimental
values of the b hadron lifetime ratios.}\label{tabl2}}\\
\vskip 0.25cm

\begin{center}
\begin{tabular}{ccc}
\hline 
&Experiment&Theory\\
\hline 
\hline 
$\frac{\tau(B^{+})}{\tau(B_{d})}$&$1.065\pm 0.023$&$1+0.05\, \left(\frac{f_{B}}{200\, 
{\rm MeV}}\right)^{2}$\\
$\frac{\tau(B_{s})}{\tau(B_{d})}$&$0.937\pm 0.040$&$1\pm 0.01$\\
$\frac{\tau({\rm b\; baryon})}{\tau(B_{d})}$&$0.773\pm 0.036$&$0.9$\\
\hline 
\hline 
\end{tabular}
\end{center}
\end{table}

A theoretical study \cite{ref6} of the lifetime problem in the light-front 
quark model suggests that the Fermi motion of the b quark inside $\Lambda_{b}$
can produce a reduction of about $12\pm 2\%$, accounting for a significant 
fraction of the discrepancy.

The measurement of the ratio of meson lifetimes has also been considered 
recently \cite{ref7} by SLD, with the result 
$\tau(B^{+})/\tau(B_{d})=1.037\pm 0.04$, to be compared with the value in 
Table \ref{tabl2}.

The Hyperon Working Group of the KTeV Collaboration at Fermilab has studied 
\cite{ref8} the $\Xi^{0}$ beta decay branching ratio
\begin{eqnarray}
\Xi^{0}\, \longrightarrow &\Sigma^{+}& +\, e^{-}\, +\, \bar{\nu}_{e}\nonumber\\
&\hookrightarrow& p\, +\, \pi^{0}
\end{eqnarray}
with a signal of $626\pm 25$ events and a background of $45\pm 18$ events. 
The process is described by the hadronic vertex of Fig.~2
\begin{figure}[!h]
\includegraphics[bb=2.75cm 5cm 17.5cm 18cm,scale=0.45]{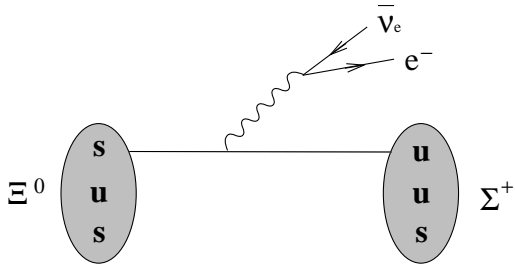}	
\vskip -2cm \caption{Hadronic vertex for $\Xi^{0}\rightarrow \Sigma^{+}\, e^{-}\,
\bar{\nu}_{e}$.}
\end{figure}
with 6 form factors, 3 vector and 3 axial. The pseudotensor (or weak electricity) 
form factor cannot be generated in the standard model with quark constituents.
The scalar and pseudoscalar form factors give contributions proportional to 
the electron mass and thus negligible. This argument is not valid for muons. 
The Collaboration aims for the extraction of the three forms: vector, magnetic 
and axial, for $\Xi^{0}$ beta decay with $2000$ events. The present result for 
the branching ratio is 
\be
BR(e)\, =\, (\, 2.60\pm\, 0.11\, \pm\, 0.16\, )\, \times\, 10^{-4}
\ee
to be compared to the theoretical $SU(3)_{f}$ predicted value 
$(2.61\pm 0.11)\times 10^{-4}$. In the CM of $\Sigma^{+}$, and using the 
$98\%$ analyzing power of $\Sigma^{+}\rightarrow p\, \pi^{0}$, 
the angular correlation between $p$ and $e^{-}$ is the decay asymmetry. 
For the muonic channel, with a few events, the measured branching ratio is
\be
BR(\mu)\, =\, (\, 3.5\, \err{2}{1}\, \err{0.5}{1}\, )\, \times\, 10^{-6}
\ee

The KTeV Hyperon Program also includes the measurement \cite{ref9} of the Hyperon 
Radiative Decays. The 1997 run has emphasized the channel 
$\Xi^{0}\rightarrow \Sigma^{0}\, +\, \gamma$, with a preliminary result 
$B.R. = ( 3.34\pm 0.12 ) \times 10^{-3}$. To obtain the asymmetry parameter, 
one must study a three stage process:
\be
\Xi^{0}\rightarrow \Sigma^{0}\, +\, \gamma\, ,\,
\Sigma^{0}\rightarrow \Lambda\, +\, \gamma\, ,\, 
\Lambda\rightarrow p\, +\, \pi^{-}
\ee
The detected particles are $p$,$\pi$ from the $\Lambda$ decay, a $\gamma$ from 
the $\Sigma^{0}$ decay and a $\gamma$ from the $\Xi^{0}$ decay. The present value 
for the asymmetry is 
\be
\alpha\, =\, -0.65\, \pm\, 0.13
\ee
More data is expected from the 1999 run.

The  theoretical studies of $\Xi^{0}\rightarrow \Sigma^{0}\, \gamma$ are based 
on the quark diagrams in Fig.~3 corresponding to the penguin diagrams 
($s\rightarrow d$ FCNC transition) plus the exchange diagram.
\begin{figure}[!h]
\includegraphics[bb=0.75cm 5cm 17.5cm 18cm,scale=0.38]{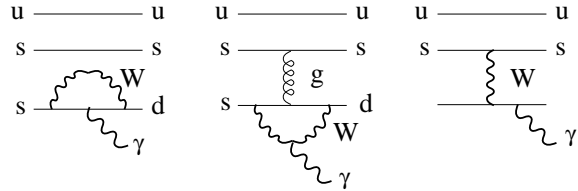}	
\vskip -2cm \caption{Quark diagrams for $\Xi^{0}\rightarrow \Sigma^{0}\, \gamma$.}
\end{figure}
The last amplitude occurs for hyperons containing a u-quark.

The radiative $\Xi^{0}$ decays, in the modes $\Xi^{0}\rightarrow \Lambda\, \gamma$ and 
$\Xi^{0}\rightarrow \Sigma^{0}\, \gamma$, have also been considered by the 
NA48 Collaboration \cite{ref10}. NA48, designed to measure $\epsilon'/\epsilon$, 
has two beam lines to generate $K_{S}$ and $K_{L}$ simultaneously and obtains 
the neutral hyperons from the $K_{S}$-Target. The results from 1997 Data are,
in units of $10^{-3}$,
\begin{eqnarray}
BR(\Xi^{0}\rightarrow \Lambda\gamma) &=& ( 1.9\pm 0.34\pm 0.19 )\nonumber\\
BR(\Xi^{0}\rightarrow \Sigma^{0}\gamma) &=& ( 3.14\pm 0.76\pm 0.32 )
\end{eqnarray}
They can be measured with $\sim 5\%$ accuracy. In 2002, the high intensity 
$K_{S}$ run will produce a statistical gain by a factor of at least 
$\sim 100$. NA48 has also found about $60$ events of the semileptonic Beta 
Decay of $\Xi^{0}$.

The future NA48 programs for $K_{S}$ and Hyperon rare decays have been 
discussed by Fantechi \cite{ref11}, in competition with KLOE and KTeV, 
respectively. A novelty is the aim to look for direct CP violation by means 
of an asymmetry in the Dalitz plot density of the three-body decays 
$K^{\pm}\rightarrow \pi^{\pm} \pi^{+} \pi^{-}$ and 
$K^{\pm}\rightarrow \pi^{\pm} \pi^{0} \pi^{0}$. Sensitivities of the order 
$10^{-4}$ are envisaged for the later phase.

An exotic role of the hyperons in Astrophysics has been presented by 
Miralles \cite{ref12}. The presence of hyperons allows a scenario in which a 
proto-neutron star is formed and emits neutrinos during tens of seconds. 
After deleptonization, it collapses to a black hole. This mechanism can be 
invoked to explain the lack of a neutron star remnant in the SN1987A and the 
detection of neutrinos from the supernova explosion.

\section{$K$-physics.}
The present world average for $\epsilon'/\epsilon$ has been discussed \cite{ref13} 
by Unal, from the NA48 Collaboration. With the results of the last two years, 
KTeV and NA48 have confirmed the original $3.5 \sigma$ finding by NA31 of 
direct CP-violation in the $K^{0}-\bar{K}^{0}$ system. With indirect 
CP-violation, i.e., in the $\Delta S=2$ mixing, established since 1964, 
the mass eigenstates $K_{S,L}$ are not pure CP eigenstates ($K_{\pm}$):
\begin{eqnarray}
K_{S} &\approx& K_{+}\, +\, \epsilon\, K_{-}\nonumber\\
K_{L} &\approx& K_{-}\, +\, \epsilon\, K_{+}
\end{eqnarray}
where $|\epsilon|\, =\, (2.28\, \pm\, 0.02) \times 10^{-3}$.

To generate Direct CP-violation, i.e., in the decay amplitude 
$|A(K^{0}\rightarrow f \bar{f})|\neq |A(\bar{K}^{0}\rightarrow f \bar{f})|$, 
one needs the interference of two decay amplitudes. The final state of two 
pions has contributions from isospin $I=0$ and $I=2$, $A_{0}$ and $A_{2}$. The 
imaginary part of the interference generated by weak CP phases (besides the 
strong phases) leads to the $\epsilon'$ parameter
\be
\epsilon'\, =\, \frac{i}{\sqrt{2}}\, {\rm Im}\left(\, \frac{A_{2}}{A_{0}}\right)\, 
e^{i(\delta_{2}-\delta_{0})}
\ee

The ratio of amplitudes from $K_{L}$ and $K_{S}$ has contributions from 
$\epsilon$ and $\epsilon'$
\begin{eqnarray}
\frac{A(K_{L}\rightarrow \pi^{+}\pi^{-})}{A(K_{S}\rightarrow \pi^{+}\pi^{-})}
\, \equiv\, \eta^{+-}\, =\, \epsilon\, +\, \epsilon'\nonumber\\
\frac{A(K_{L}\rightarrow \pi^{0}\pi^{0})}{A(K_{S}\rightarrow \pi^{0}\pi^{0})}
\, \equiv\, \eta^{00}\, =\, \epsilon\, -\, 2 \epsilon'
\end{eqnarray}
In order to separate $\epsilon'$ experimentally, one considers the ratio of 
ratios of decay rates
\begin{eqnarray}
R &=& \frac{\Gamma(K_{L}\rightarrow \pi^{0}\pi^{0})\,\Gamma(K_{S}\rightarrow \pi^{+}\pi^{-})}
{\Gamma(K_{S}\rightarrow \pi^{0}\pi^{0})\,\Gamma(K_{L}\rightarrow \pi^{+}\pi^{-})}
\nonumber\\
&=& 1\, -\, 6\, {\rm Re}\left(\frac{\epsilon'}{\epsilon}\right)
\end{eqnarray}
To establish Direct CP Violation, one needs $R\neq 1$.

In the standard model, $\epsilon$ is generated from the box diagram whereas 
$\epsilon'$ gets its dominant value from gluonic and electroweak penguin 
diagrams.

The experimental situation is pictured in the Figure 4.
\begin{figure}[!h]
\includegraphics[bb=1.5cm 5cm 17.5cm 18cm,scale=0.38]{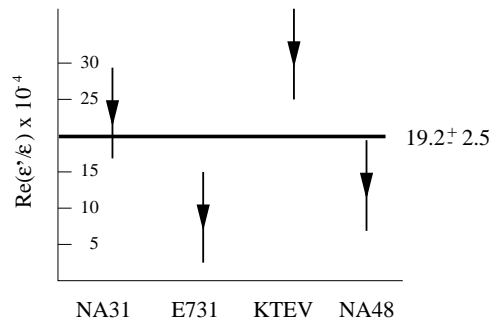}	
\vskip -2cm \caption{Experimental measurements of $\epsilon'/\epsilon$.}
\end{figure}

One realizes from the world average value that $\epsilon'/\epsilon\neq 0$ is 
well established, but the actual value is probably not.
As illustrated in Figure 4, the $\chi^{2}$ is poor. More results from NA48, 
KTeV and KLOE, which uses a different method, will clarify the experimental 
situation.

The establishment of Direct CP-violation  tells us that a superweak \cite{ref14} 
explanation is ruled out, and that the K-system needs a milliweak model to 
describe CP violation. In the standard model, this description is understood 
in terms of the relative magnitudes of the sides of the unitary triangle, 
shown in Fig 5.
\begin{figure}[!h]
\includegraphics[bb=2.5cm 5cm 17.5cm 18cm,scale=0.45]{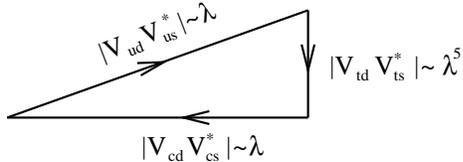}	
\vskip -2.75cm \caption{The $(sd)$ unitarity triangle.}
\end{figure}
The CP-asymmetries are thus expected to be of order $\lambda^{4}$.

The calculation of the two isospin amplitudes $A_{0,2}$ makes use of the 
$\Delta S=1$ effective hamiltonian, which leads to four-quark operators of the 
current-current form ($Q_{1,2}$), QCD penguin ($Q_{3\rightarrow 6}$) and 
electroweak penguin ($Q_{7\rightarrow 10}$) diagrams. In conventional notation, 
$Q_{6}$ and $Q_{8}$ are most important, but their matrix elements have opposite 
signs. Possible cancellations are thus a potential danger in the theoretical 
calculations. In fact, ${\rm Im}\, A_{0}$ is dominated by $Q_{6}$ whereas 
${\rm Im}\, A_{2}$ is dominated by $Q_{8}$.

It is well known that, around $500$ MeV, the $\pi\, \pi$ interaction is very 
strong in the scalar-isoscalar channel. The role of final state interactions 
is thus very important \cite{ref15} for the $A_{0}$-amplitude. What is a subject 
of debate \cite{ref16} is whether the dispersive computation has enough 
reliability. This is a difficult problem, but the Omn\`es resummation of chiral 
logarithms \cite{ref15} gives a $50\%$ enhancement. In this case, $A_{0}$ and thus 
$Q_{6}$, is the primary problem. This leads to an estimate $\epsilon'/\epsilon =
(15\pm 5)\times 10^{-4}$, compatible with the present experimental value.

Models of low energy dynamics point towards a connection between the 
$\Delta I=1/2$ rule and a "large" $\epsilon'/\epsilon$. 
Lattice QCD simulations will tell us whether this suggestion has a firm ground.

A review on the final CPLEAR results on CP, T and CPT in the neutral kaon 
system was presented by Zavrtanik \cite{ref17}. For the $2\pi$ decay channel, 
this experiment has observed for the first time a difference in the time 
dependence of the $K^{0}$ and $\bar{K}^{0}$ decay rates.
CP violation implies T violation or CPT violation 
  or both. Is T-violated? CPLEAR has measured the Kabir asymmetry \cite{ref18}, 
  by comparing $K^{0}\rightarrow \bar{K}^{0}$ versus $\bar{K}^{0}\rightarrow K^{0}$.
The flavour tag at the production time is defined by the charged 
kaon $p \bar{p}\rightarrow K^{-} \pi^{+} K^{0}$, $K^{+} \pi^{-}
\bar{K}^{0}$. The strangeness of the neutral kaon at the decay time is 
defined by the lepton charge in the semileptonic decay ($\Delta S=\Delta Q$).
The Kabir asymmetry is a genuine T-violating observable, which needs both 
T-violation and $\Delta \Gamma\neq 0$. This method works thus for neutral kaons, 
due to the difference in $K_{S}$ and $K_{L}$ lifetimes. CPLEAR results are 
compatible with equal CP and T violations and CPT invariance.

\section{Charm and Beauty production}
In the production of heavy hadrons, the fragmentation function \cite{ref19} is 
the link between the heavy quark and the heavy hadron. It is parametrized by 
the probability $f(z)$ that a hadron shares a fraction $z$ of the quark momentum
\be
z\, =\, \frac{( E\, +\, p_{\Vert} )_{H}}{( E\, +\, p )_{Q}}
\ee
The problem with the variable $z$ is the denominator, which refers to the 
quark before fragmentation, so that $z$ is not accessible on a 
event-by-event basis. New variables which are experimentally accessible 
are defined as the hadron energy with respect to the beam energy
\be
x_{E}\, \equiv\, \frac{E_{H}}{E_{beam}}
\ee
There are recent results on $\langle x_{B} \rangle$ for the B meson from 
ALEPH, DELPHI, OPAL and from SLD, with values for the leading B energy ranging 
from $0.72\rightarrow 0.74$. 
The methodology follows different strategies, and still one has to 
understand the consistency of the different analyses. At SLD, the 
polarization of the electron beam is used to tag b-quarks with $100\%$ 
efficiency \cite{ref7}. The reconstruction of the secondary vertex by exploiting 
the kinematics leads to a measurement of the mean energy of weakly-decaying
B hadrons, with a value $\langle x_{B} \rangle\, =\, 0.714\pm 0.009$.

The SELEX experiment \cite{ref20} emphasizes the understanding of charm 
production in the forward hemisphere. QCD factorization predicts that heavy 
quarks hadronize through jet fragmentation functions independently of the 
initial state. The experimental data show that the produced charm (anticharm) quark
combines with a projectile valence quark. $\Lambda^{+}_{c}$ is a leading 
particle when produced by the 3 beams $\pi^{-}$, $p$, $\Sigma^{-}$. 
The $\Lambda_{c}$ hadroproduction has a hard $x_{F}$ distribution. 
There is a strong production asymmetry in favour of $\Lambda^{+}_{c}$ over 
$\Lambda^{-}_{c}$ for baryon beams. It is less strong for a $\pi^{-}$ beam.

\section{Charm and Beauty decays}
The semileptonic b-decay studies have the double objective of the 
understanding of the dynamics of heavy quark decays plus the extraction 
of the CKM coupling constants $V_{cb}$, $V_{ub}$. The inclusive $BR(b)_{SL}$ 
has been discussed by Margoni \cite{ref21}, with different strategies of 
b-lifetime and lepton tags exploited at LEP. For the first time, DELPHI has 
explicitly separated by direct measurement $BR(b\rightarrow \bar{c}\rightarrow
l^{-})$. At present the analyses to 
extract $V_{cb}$ are mainly limited by theory ($b\rightarrow l$, 
$b\rightarrow c\rightarrow l$ decay models). The most 
precise determination in the OPE approach to analyze the LEP data gives
\begin{eqnarray}\label{eq16}
|V_{cb}|^{\rm incl}_{\rm LEP} &=& (\, 40.76 \pm 0.41\, (\mbox{\rm exp.})\nonumber\\
&\pm&  2.04\, (\mbox{\rm theo})\, )\times 10^{-3}
\end{eqnarray}

The alternative to the inclusive decay is the study of the exclusive 
$B^{0}\rightarrow D^{*}\, l\, \nu$ decay as a function of the $D^{*}$ recoil
\be
\frac{d \Gamma}{d \omega}\, =\, K(\omega)\, F^{2}(\omega)\, |V_{cb}|^{2}
\ee
where $K(\omega)$ is a phase space factor and $F(\omega)$ is the Isgur-Wise 
form factor, for which the heavy-quark-effective-theory value for no recoil 
$\omega=1$ is estimated. The problem is that, due to $K(\omega)$, the decay 
rate vanishes at $\omega=1$. The procedure is thus the measurement of 
$\frac{d \Gamma}{d \omega}$ to fit it and extrapolate 
 to $\omega=1$ to obtain $F(1)\, |V_{cb}|$. These measurements at LEP have been 
 presented by Terem \cite{ref22}. There is a problem with $b\rightarrow D^{**}\, 
 l\, \nu$, followed by $D^{**}\rightarrow D^{*+}\, X$, which is 
 an important systematic effect. ALEPH and DELPHI fit to $D^{*}$ and $D^{**}$  
 contributions gives
\begin{eqnarray}
\lefteqn{\mbox{\rm Br}(B^{-}\rightarrow D^{**0}\left( \rightarrow D^{*+} \pi^{-}\right)\,
l\, \nu )}\nonumber\\
&=& (\, 1.24\, \pm 0.19 \pm 0.04\, )\, \%
\end{eqnarray}
The LEP average for the exclusive analysis gives 
$|F(1) V_{cb}|\, =\, ( 34.9\, \pm 1.7 )\times 10^{-3}$, with 
higher experimental error than (\ref{eq16}) due to small samples, but 
much cleaner theoretical approach.

The determinations of $V_{ub}$ can come from either the exclusive 
$B\rightarrow \pi, \rho\, l\, \nu$ decays, 
where the main limitation is statistics or the inclusive lepton endpoint 
analysis, above the process $b\rightarrow c\, l\, \nu$. This method, 
limited by theoretical uncertainties, extracts \cite{ref23} a measurement of 
the branching ratio for 
inclusive charmless semi-leptonic b decays
\be
\mbox{\rm Br}(b\rightarrow X_{u}\, l\, \nu )\, =\, (\, 1.67\, \pm 0.60\, )\times
10^{-3}
\ee
from ALEPH, DELPHI and L3 at LEP. The LEP average for the $V_{ub}$ value 
derived using HQET is
\be
|V_{ub}|\, =\, \left(\, 4.04\, \err{0.62}{0.74}\, \right)\times 10^{-3}
\ee

ALEPH takes 4 million $e^{+}\, e^{-}\rightarrow Z\rightarrow q\, \bar{q}$ 
events to measure \cite{ref24} the branching 
fractions for $D_{s}\rightarrow \tau\, \nu$ ($\tau\rightarrow e\, \nu \bar{\nu}$
or $\tau\rightarrow \mu\, \nu\, \bar{\nu}$) and $D_{s}\rightarrow \mu\, \nu$. 
Due to chirality suppression in the pseudoscalar decay of Fig. 6
\begin{figure}[!h]
\includegraphics[bb=2.5cm 5cm 17.5cm 18cm,scale=0.45]{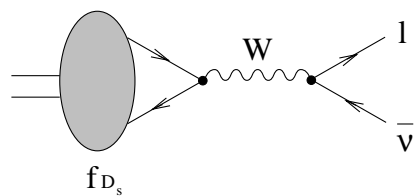}	
\vskip -2.75cm \caption{Pseudoscalar decay $D_{s}\rightarrow l\, \bar{\nu}$.}
\end{figure}
the $D_{s}\rightarrow e\, \nu$ decay is not accessible. The two leptonic
$\tau$ decay channels, which give consistent signals, measure
\be\label{eq21}
\mbox{\rm Br}(D_{s}\rightarrow \tau\, \nu )\, =\, (\, 5.79\, \pm 0.76\pm 1.78\, ) \%
\ee
whereas the $D_{s}\rightarrow \mu\, \nu$ analysis gives
\be\label{eq22}
\mbox{\rm Br}(D_{s}\rightarrow \mu\, \nu )\, =\, (\, 0.68\, \pm 0.11\pm 0.18\, ) \%
\ee
The two results (\ref{eq21}) and (\ref{eq22}) are consistent with the chirality 
suppression and phase space factors $m_{l}^{2}\, \left(\, 1\, -\, 
\frac{m_{l}^{2}}{M_{D_{s}}^{2}}\right)^{2}$ and provide a proof of 
leptonic universality in charged current decays. Combinig them, one gets 
for the decay constant
\be
f_{D_{s}}\, =\, (\, 285\pm 20\pm 40\, )\, \mbox{\rm MeV}
\ee
which can be used to check the validity of its prediction by different 
theoretical models. Lattice QCD predicts a value $240\err{30}{25}$ MeV.

Hadronic decays $B^{*}\rightarrow B\, \pi$, $D^{*}\rightarrow D\, \pi$ 
allow \cite{ref25} the extraction of the 
physical coupling
\begin{eqnarray}
\langle B^{0}(p)\, \pi^{+}(q) | B^{*+}(p') \rangle &=& g_{B^{*} B \pi}(q^{2})\,
\epsilon^{\mu}\, q_{\mu}\\
g_{B} &=& \lim_{q^{2}\rightarrow m_{\pi}^{2}} g_{B^{*} B \pi}(q^{2})\nonumber
\end{eqnarray}
\begin{figure}[!h]
\includegraphics[bb=4.5cm 7cm 17.5cm 16cm,scale=0.60]{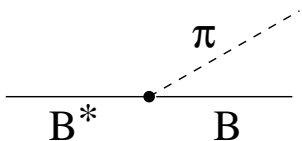}	
\vskip -3.75cm \caption{Vertex $B^{*}$-$B$-$\pi$.}
\end{figure}
and analogously for the $D^{*}\, D$ transition.
These form factors have been obtained theoretically from QCD sum rules 
\cite{ref26}. They cannot be described by a monopole function. Two different 
methods give consistent results and one gets for the coupling constants
\be
g_{D}\, =\, 5.7\pm 0.4\;\;\;\;\;\;,\;\;\;\;\;\; g_{B}\, =\, 14.5\pm 3.9
\ee

The excited states of $D$, $D_{s}$, $B$ and $B_{s}$ mesons have been studied 
\cite{ref27} in the framework of the relativistic heavy chiral quark model, 
with the determination of spectrum and wavefunctions. The $1/m_{h}$-effects 
are relevant for the calculation of the decay amplitudes of $B^{**}\rightarrow
B+\eta$, $\pi$ and $K$. Decay channels of 
$B^{**}$ are useful for flavour tagging in particle detectors.

The decays $D^{+}$, $D^{+}_{s}\rightarrow \pi^{-} \pi^{+} \pi^{+}$ were studied 
experimentally by the E791 Collaboration 
\cite{ref28}, at Fermilab fixed target programme. The experiment runs for 
$500$ GeV $\pi^{-}$-nucleon interactions and the signals yield 
($1240\pm 51$) $D^{+}$ events and ($858\pm 49$) $D^{+}_{s}$ events. 
Besides the branching 
ratios, a detailed analysis of the Dalitz 
plots has been perfomed. The invariant mass $M_{\pi^{+}\pi^{-}}^{2}$ 
distribution for $D^{+}_{s}\rightarrow \pi^{-} \pi^{+} \pi^{+}$
is completely dominated by $f_{0}(980)$ and $f_{0}(1370)$, as shown in the 
Fig.~8, with a negligible non-resonant contribution.
\begin{figure}[!h]
\includegraphics[bb=2.5cm 2.5cm 17.5cm 21cm,scale=0.45]{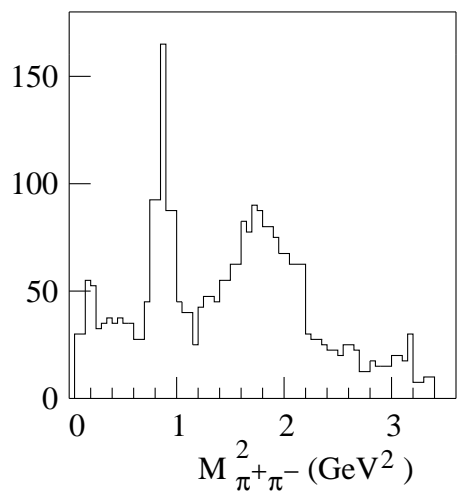}	
\vskip -2.75cm \caption{$M_{\pi^{+}\pi^{-}}^{2}$ 
distribution for $D^{+}_{s}\rightarrow \pi^{-} \pi^{+} \pi^{+}$.}
\end{figure}
The behaviour of the $M_{\pi^{+}\pi^{-}}^{2}$ distribution for
$D^{+}\rightarrow \pi^{-} \pi^{+} \pi^{+}$ is however completely 
different, with a dominant non-resonant contribution shown in Fig.~9.
\begin{figure}[!h]
\includegraphics[bb=2.5cm 3cm 17.5cm 18cm,scale=0.45]{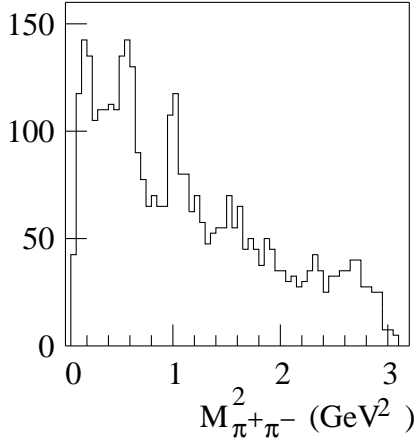}	
\vskip -2.75cm \caption{$M_{\pi^{+}\pi^{-}}^{2}$ 
distribution for $D^{+}\rightarrow \pi^{-} \pi^{+} \pi^{+}$.}
\end{figure}
What is the origin of the low mass peak? One is led naturally to the 
$\sigma$-meson, the scalar-isoscalar predicted by Nambu and Jona-Lasinio in a 
linear realization of the chiral Lagrangian. Experimentally, it has 
suffered all kinds of up's and down's in the Review of particle properties 
along the years. The inclusion of the $\sigma$ in the fit leads to an 
spectacular improvement and to the determination of its mass ($483\pm 30$) MeV
and width ($338\pm 50$) MeV. The light $\sigma(500)$, in spite of its broadness,
is up again !

Rare decays are a good probe for searching new physics. The branching 
ratios for the inclusive $B\rightarrow X_{s}\, l^{+}\, l^{-}$ and exclusive
$B\rightarrow K^{(*)}\, l^{+}\, l^{-}$ decays are smaller in 
the standard model than the experimental bounds, so that there is room for 
contributions from models beyond the standard theory. In particular, there 
is a very interesting property \cite{ref29} in 
$B\rightarrow K^{*}\, l^{+}\, l^{-}$: the zero of the 
forward-backward asymmetry provides a discrimination between the standard 
model and supersymmetry.

The very rare $\Delta S=2$ process $b\rightarrow s\, s\, \bar{d}$ is described 
in the standard model by the box diagram of Fig.~10.
\begin{figure}[!h]
\includegraphics[bb=2.5cm 5cm 17.5cm 18cm,scale=0.45]{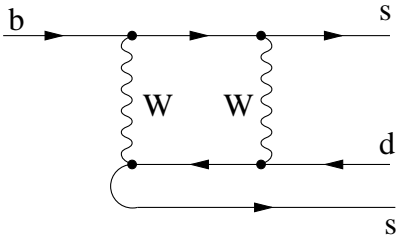}	
\vskip -2.75cm \caption{Standard Model box diagram for $b\rightarrow s s \bar{d}$.}
\end{figure}
with a branching ratio of the order $10^{-11}$. Whereas the MSSM squark-gaugino 
box diagram in Fig.~11
\begin{figure}[!h]
\includegraphics[bb=2.5cm 5cm 17.5cm 18cm,scale=0.45]{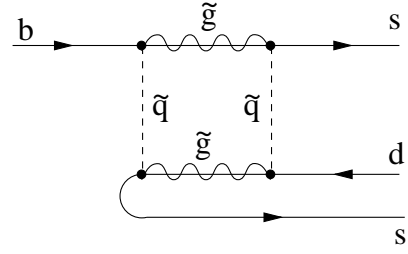}	
\vskip -2.75cm \caption{MSSM squark-gaugino box diagram for 
$b\rightarrow s s \bar{d}$.}
\end{figure}
can increase \cite{ref30} the theoretical branching ratio to levels of $10^{-8}$, 
the MSSM with $R$ parity violating couplings induced by the sneutrino, 
see Fig.~12, is not 
restricted.
\begin{figure}[!h]
\includegraphics[bb=2.5cm 5cm 17.5cm 18cm,scale=0.45]{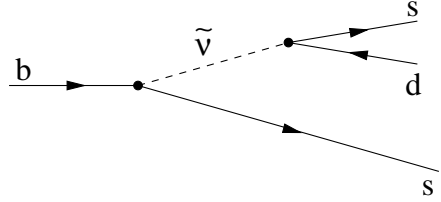}	
\vskip -2.75cm \caption{MSSM sneutrino-mediated diagram for 
$b\rightarrow s s \bar{d}$.}
\end{figure}
On the contrary, data from LEP1, around the $Z$ resonance, allow the 
search for $B^{-}\rightarrow K^{-}\, K^{-}\, \pi^{+}$ \cite{ref31}. 
The upper limit of $\sim 10^{-4}$ for the branching ratio 
leads to new limits on the contribution of $R$ parity violating couplings 
in this process.

\section{B(D) mixing and CP-violation}
For charm mesons, the two parameters of mixing
\be
x\, =\, \frac{\Delta M}{\Gamma}\;\;\;\;\;\;, \;\;\;\;\;\; y\, =\, 
\frac{\Delta \Gamma}{2\, \Gamma}
\ee
are small. The experimental methods to see $x$ or $y$ are either by mixing, 
with wrong sign final lepton, or comparing the lifetime of CP eigenstates. 
The last method takes into account the expectation that CP-violation for 
the charm sector is small. FOCUS(E831) selects \cite{ref32} the two channels
\begin{eqnarray}
D^{0} &\rightarrow& K^{+}\, K^{-}\, (\mbox{CP}\, + )\nonumber\\
D^{0} &\rightarrow& K^{-}\, \pi^{+}\, (\mbox{CP}\, +\, :\, \mbox{CP}\, -\, =\, 1\, :\, 1 )
\end{eqnarray}
and the direct comparison of CP final state lifetimes finds $y_{\rm CP}$ as
\be
y_{\rm CP}\, =\, \frac{\tau(D\rightarrow K\, \pi)}{\tau(D \rightarrow K\, K)}\, -\, 1
\ee
The experimental result is $(3.42\pm 1.39\pm 0.74)\%$.

The standard model predicts that direct CP violation in D decay rates is 
the largest in singly Cabbibo-suppressed decays $D^{+}\rightarrow K^{-}\,
K^{+}\, \pi^{+}$, $D^{0}\rightarrow K^{-}\, K^{+}$, $\pi^{-}\, \pi^{+}$.
The CP asymmetry results \cite{ref32} show no evidence for CP violation at the 
 level of few percent.
 
In the $b$ sector, the problems of $B_{d}$ and $B_{s}$ mixing allow the extraction 
of $V_{td}$ and $V_{td}/V_{ts}$ matrix elements of the CKM matrix. In the time 
integrated  approach, the $B_{d}$-mixing leads to a world average 
$\Delta m_{d} = 0.484\pm 0.015$ ps$^{-1}$, but 
the $B_{s}$-mixing has no sensitivity to $\Delta m_{s}$.
The method used \cite{ref33,ref37} needs a time dependent experimental approach. 
The time dependent mixing generates a periodic signal. The amplitude fit method
measures the oscillation amplitude $A$ at fixed frequency $\Delta m_{s}$. One 
expects $A=1$ on a frequency equal to the true $\Delta m_{s}$, whereas 
$A=0$ for a wrong frequency. The world combination leads to the conclusion that 
$B_{s}$ oscillations have not yet been resolved, with a lower limit 
$\Delta m_{s}> 14.6$ ps$^{-1}$. The standard model preferred value is close to 
the present reach. In the ALEPH data, there is a hint of a signal around 
$17$ ps$^{-1}$. With expected sensitivities like $\sim 19$ ps$^{-1}$, one can 
envisage very interesting results in the near future.

The measurement of $\Delta \Gamma_{s}$ has been addressed with many methods 
\cite{ref5}. An appreciable value would allow to see CP violation in untagged 
$B_{s}$, contrary to $B_{d}$ in which $\Delta \Gamma_{d}\approx 0$. With the 
constraint $\Gamma_{s}\, =\, \Gamma_{d}\, =\, \Gamma$, the present combined 
experimental value by the LEP Working Group is 
$\frac{\Delta \Gamma_{s}}{\Gamma} = 0.16\err{0.08}{0.09}$.
This is still an insufficient sensitivity to claim an observed width difference.
The standard model preferred value is $0.05\pm 0.03$ \cite{ref34}, using 
Lattice HQET and extrapolated Lattice QCD.

One of the highlights of the Conference is the presentation that the two 
B-factories and the corresponding detectors, BABAR at PEP-II 
\cite{ref35,ref36} and BELLE at KEK B \cite{ref37,ref38}, are 
working very well. The PEP-II $9$ GeV $e^{-}$ against $3.1$ GeV
$e^{+}$ collider expects a 
luminosity of about $6\times 10^{33}$ cm$^{-2}$ s$^{-1}$ around summer 2000, 
whereas the KEK B $8$ GeV $e^{-}$ against $3.5$ GeV $e^{+}$ collider had a 
luminosity about $2\times 10^{33}$ cm$^{-2}$ s$^{-1}$ just  
before the Conference in June 2000. Some of the many physics results which 
are being analyzed by the two Collaborations have been presented, in 
particular, the lifetimes $\tau(b)$, $\tau(c)$ and $\tau(\tau)$, the mixing 
$\Delta m_{d}$, the inclusive $B\rightarrow J/\psi\, X$ decay, the charmless
$B\rightarrow \rho\, \pi$ decay, the dominant "Cabibbo allowed" 
$\mbox{\rm Br}(B\rightarrow D^{*}\, \pi)$ 
and $\mbox{\rm Br}(B\rightarrow D_{s}^{*+}\, D^{*-})$, or the rare 
$B\rightarrow K^{*}\, \gamma$, $B\rightarrow K^{*}\, l^{+}\, l^{-}$  decays.

The main objective of the B-factories is to establish CP-violation outside 
the K-system and see whether its description obeys to the CKM mixing matrix 
in the standard model. For the $(bd)$ unitarity triangle, the three sides mediated
by $u$, $c$, $t$ quarks are of similar size of order $\lambda^{3}$. When the CP 
conserving direction is taken as a reference and the associated side is 
normalized to one, the $(bd)$ unitarity triangle is shown in Fig.~13.
\begin{figure}[!h]
\includegraphics[bb=2.5cm 5cm 17.5cm 18cm,scale=0.45]{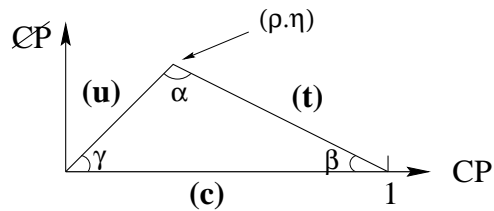}	
\vskip -2.75cm \caption{$(bd)$ unitarity triangle.}
\end{figure}
B-Physics has the power to find observables able to overconstrain the 
parameters of this triangle. The separate measurement of the weak CP phases 
$\alpha$, $\beta$, $\gamma$ is possible. The value of $\sin(2\beta)$ is 
accessible from the CP asymmetry 
in $B\rightarrow J/\psi\, K_{s}$ generated from the interplay of mixing and 
decay. This method needs a flavour tag, as given by the lepton channel or 
others. Suppose the decay of $\Upsilon(4S)$ into an entangled state of two 
$B$'s:
\be
\Upsilon(4S)\rightarrow B_{1}\, B_{2}\, \left\{
\begin{array}{l}
\Rnode{a}{\bullet}\;\;\;\;\;\; \Rnode{b}{\phantom{l}}B\rightarrow D^{-}\, e^{+}\,\nu\, \mbox{\rm (Tag)}\\
\\[3mm]
\Rnode{c}{\bullet}\;\;\;\;\;\;\;\;\;\;\;\;\Rnode{d}{\phantom{l}}\bar{B}\rightarrow
J/\psi\, K_{s}\\[-2mm]
\phantom{\bullet\;\;\;\;\;}\Rnode{e}{\;}\;\;\;\;\;\Rnode{f}{\;}
\end{array}
\right.
\psset{arrows=->}
\ncLine{a}{b}
\ncLine{c}{d}
\psset{arrows=-}
\ncline[linestyle=none]{e}{f}\mput{\scriptstyle \Delta z}
\ee
After a time $\Delta t$ (or length $\Delta z$) from the tag, the CP eigenstate 
$B_{-}$ is 
observed: the comparison of $B^{0}\rightarrow B_{-}$  versus 
$\bar{B}^{0}\rightarrow B_{-}$ measures CP violation.

Ba\~nuls \cite{ref39} has discussed the way to use these decays to look for T and CPT 
violation. Starting from the transition $B^{0}\rightarrow B_{-}$, one has
\be
B^{0}\rightarrow B_{-}\, \Rnode{a}{\phantom{aaaaa}}\begin{array}{l}
\Rnode{b}{\phantom{A}}B_{-}\, \rightarrow\, B^{0}\\
\\
\\
\Rnode{c}{\phantom{B}}B_{-}\, \rightarrow\, \bar{B}^{0}
\end{array}
\psset{arrows=->}
\ncLine[doubleline=true]{a}{b}\Aput{T}
\ncLine[doubleline=true]{a}{c}\Bput{CPT}
\ee
These transformed transitions need a CP tag. In order to project first on 
$B_{-}$, the $B$ of the other side has to be identified as $B_{+}$, a difficult 
problem. There is, however, an equivalent transition for hermitian 
hamiltonians. In the limit $\Delta\Gamma_{d}=0$ (an excellent approximation 
for $B_{d}$), the transition $B_{-}\rightarrow B^{0}$ is equivalent to 
$B_{+}\rightarrow \bar{B}^{0}$, which is obtained from the original 
$B^{0}\rightarrow B_{-}$ by a temporal exchange $\Delta z\rightarrow -\Delta z$
of the two decay channels: leptonic and $J/\psi\, K_{s}$. Although the temporal
and T-odd asymmetries are conceptually different, they 
become equivalent in the limit $\Delta\Gamma_{d}=0$ and the temporal asymmetry 
is a T-odd observable.

The problem of B-physics and CP-violation is a source of inspiration and 
dedication in the hadronic machines too. At the Conference, the prospects of 
CDF-II \cite{ref40}, BTeV \cite{ref41}, CKM \cite{ref42}, and Run II \cite{ref43} at 
FermiLab, as well as HERA b \cite{ref44} at DESY and ATLAS \cite{ref45}, 
CMS \cite{ref46} and LHCb \cite{ref47} at LHC \cite{ref48} were given. A brilliant scenario 
appears at the near future. Contrary to the B-factory 
 preparation, the B-production mechanism is thought to be here incoherent from 
 the individual $b$ quark, with no entanglement. One can proceed then to 
 flavour tags, but there is no possibility of CP tags as discussed above.
 
It is worth to emphasize that, inside the Standard Model, the CP phases can 
be also extracted from CP conserving observables in exclusive B-decays. For 
example, $\cos \alpha$ is measured in the rare $B\rightarrow \rho\gamma$ decay 
\cite{ref49,ref50}. The evidence 
for $\alpha\neq 0$ in the $(bd)$ unitary triangle is here $\cos \alpha\neq 1$ 
in these observables, 
contrary to $\sin \alpha\neq 0$ (yes-no experiment) in the CP-odd asymmetries.

The chapter of SM Physics was also addressed by Narain \cite{ref51}, with an 
excellent review of Top Quark Physics at the Tevatron (Run I and Run II) 
and the LHC as a top factory.

\section{Heavy Quarkonium}
Quarkonia are special hadrons, for which a description in terms of 
factorization between the hard and soft scales is believed to be valid. 
For the case of $\Upsilon$, confinement effects are small and basic 
perturbative QCD calculations on some observables \cite{ref52} can be envisaged.
These observables refer to structure and decays. The only flavour dependent 
parameter is the $b$ quark mass $m$ at the scale of the bound state. 
The running of the $b$ quark mass has been established \cite{ref53}. With the 
velocity $v\sim \alpha$ for Coulomb systems, one has a multiscale problem with 
$m$, $p\sim m v$, $E\sim m v^{2}$. The separation of scales is made by means of 
an effective field theory \cite{ref54}.

There is a $J/\psi$ and $\psi'$ surplus in direct production at the Tevatron.
An interesting production mechanism which has been suggested is the 
Colour-Octet component in NRQCD. Chao \cite{ref55} has discussed a test of this
mechanism, which incorporates the colour-octet gluon fragmentation shown 
in Fig.~14
\begin{figure}[!h]
\includegraphics[bb=1.5cm 5cm 17.5cm 18cm,scale=0.45]{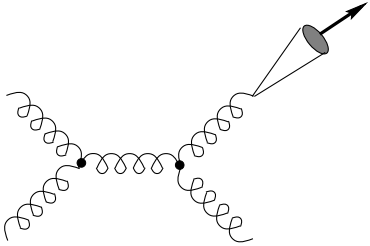}	
\vskip -2.75cm \caption{Color-octet gluon fragmentation.}
\end{figure}
based on the polarization of charmonium at the Tevatron. In the NRQCD 
factorization approach, an explicit calculation of the production cross 
section \cite{ref56} shows that the colour-octet contributions can describe
Tevatron data. The experimental $J/\psi$ polarization at high $p_{T}$ is, 
however, in disagreement with the calculation for direct $J/\psi$ production 
plus the feeddown from intermediate $\chi_{c}$ and $\psi'$. More tests are needed
to understand this problem.

The inelastic $J/\psi$ production in DIS ($2<Q^{2}< 80$ GeV$^{2}$) at HERA 
shows \cite{ref57} that the 
colour-singlet contribution is below data, but the shape is in reasonable 
agreement. When the colour-octet contribution, as suggested by the Tevatron 
data, is included, the theoretical magnitude is above data and the shapes 
disagree. Clearly one has to conclude that the problem of the production 
mechanism is not understood yet.

\section{Outlook}
The Conference was a great event. Many experimental results and theoretical 
ideas were presented and discussed. The understanding of the Flavour Problem 
is one of the main pending questions in fundamental physics. In the quark 
sector, this study involves strange, charm and beauty hadrons, and so the 
control of the interplay between electroweak and strong interactions.
It is gratifying for this field that all major facilities in particle 
physics around the world have a strong programme in it. As a consequence, 
we can expect important breakthroughs in the next two years and to have a 
fruitful rendez-vous at Vancouver 2002.

\section{Acknowledgements}
I would like to thank the Organizers of the Conference for the appropriate 
environment and to Vicent Gim\'enez for a critical reading of the manuscript.
 This work was supported by CICYT, under Grant AEN99-0692.

\end{document}